\documentclass[11pt]{article}
\usepackage{graphics}
\usepackage{psfrag}

\title{$CP$ Violation in the Decays $B \to K \overline{K}$ in the Presence of $D \overline{D} \leftrightarrow K \overline{K}$ Mixing}

\author{S. Barshay$^1$, L. M. Sehgal\footnote{{\it E-mail address}: sehgal@physik.rwth-aachen.de (L. M. Sehgal)}$\,\, ^2$  and J. van Leusen$^2$ \\ 
\it{$^1$III. Physikalisches Institut, RWTH Aachen} \\
\it{$^2$Institut f\"ur Theoretische Physik, RWTH Aachen} \\
\it{D-52056 Aachen, Germany}}

\date{}
\begin{document}

\maketitle

\begin{abstract}
We have recently shown that the large direct $CP$ violation observed in the decay $B^0 \to \pi^+ \pi^-$, and the enhanced branching
ratio for $B^0 \to \pi^0 \pi^0$, can be understood by invoking a small mixing of the $\pi \pi$ system with the dominant $D \overline{D}$ channel.
We examine here the analogous effect of $D \overline{D} \leftrightarrow K \overline{K}$ mixing on the rare decays $B^0 \to K^0 \overline{K^0}$,
$B^- \to K^- K^0$ and $B^0 \to K^+ K^-$. We find (a) significant values for the asymmetry parameters ${\cal C}$ and ${\cal S}$ in 
$B^0 / \overline{B^0} \to K^0 \overline{K^0}$, (b) a possible enhancement of the suppressed mode $B^0 \to K^+ K^-$, (c) a correlation
between the three decay channels following from a triangle relation between amplitudes $A_{K^0 \overline{K^0}} - A_{K^+ K^-} = A_{K^- K^0}$.
The pattern of asymmetries and branching ratios is compared with that derived from the short-distance QCD penguin interaction.
\end{abstract}

The rare decays $B^0 \to K^0 \overline{K^0}$, $B^- \to K^- K^0$ and $B^0 \to K^+ K^-$ have a rather simple description in the conventional
approach, based on the short-distance Hamiltonian for $b \to d s \overline{s}$ and the assumption of factorization of hadronic matrix
elements \cite{Fleischer,BSW,Neubert;Stech,Beneke;Neubert}. The dominant interaction is the QCD penguin transition 
$b \to d g^{\ast} \to d s \overline{s}$, which leads to the following amplitudes \cite{Fleischer} (the absolute square of an amplitude
gives the branching ratio):
\begin{eqnarray}
A( \overline{B^0} \to K^0 \overline{K^0} ) & = & N_0 [ \lambda_t + \Delta P \lambda_c ] \nonumber \\
A(B^- \to K^- K^0) & = & A(\overline{B^0} \to K^0 \overline{K^0}) \label{AmpFleischer}\\
A(\overline{B^0} \to K^+ K^-) & = & 0 \nonumber
\end{eqnarray}
Here $\lambda_t = V_{tb} V_{td}^{\ast}$, $\lambda_c = V_{cb} V_{cd}^{\ast}$ are CKM factors, and $\Delta P$ is a term of order $(m^2_c -m^2_u)/m^2_b$
that we discuss below. The normalization $N_0$ contains the penguin factor $\frac{ \alpha_s}{12 \pi} \ln \frac{m^2_t}{m^2_u}$, as well as
decay constants and wave-function overlaps. A typical estimate is $N_0 \approx 0.12$, which implies a branching ratio 
${\rm Br} (B^0 / \overline{B^0} \to K^0 \overline{K^0}) \approx 1.0 \times 10^{-6}$. The notable feature of Eqs. (\ref{AmpFleischer}) is the
equality of the decay rates of $\overline{B^0} \to K^0 \overline{K^0}$ and $B^- \to K^- K^0$, as well as equal $CP$-violating asymmetries
between $\overline{B^0} \to K^0 \overline{K^0}$ and $B^0 \to K^0 \overline{K^0}$, and between $B^- \to K^- K^0$ and $B^+ \to K^+ \overline{K^0}$.
Also notable is the suppression of the decay $\overline{B^0} \to K^+ K^-$ (which persists even after inclusion of annihilation contributions, the 
branching ratio being $\sim 10^{-8}$ \cite{Beneke;Neubert}).

\begin{figure}
\center
\psfrag{C00}[cc][][2]{${\cal C}$}
\psfrag{s}[cc][][1.8]{$q^2 / (GeV)^2$}
\psfrag{S00}[cc][][2]{${\cal S}$}
\makebox[7cm]{\resizebox{7cm}{!}{\includegraphics{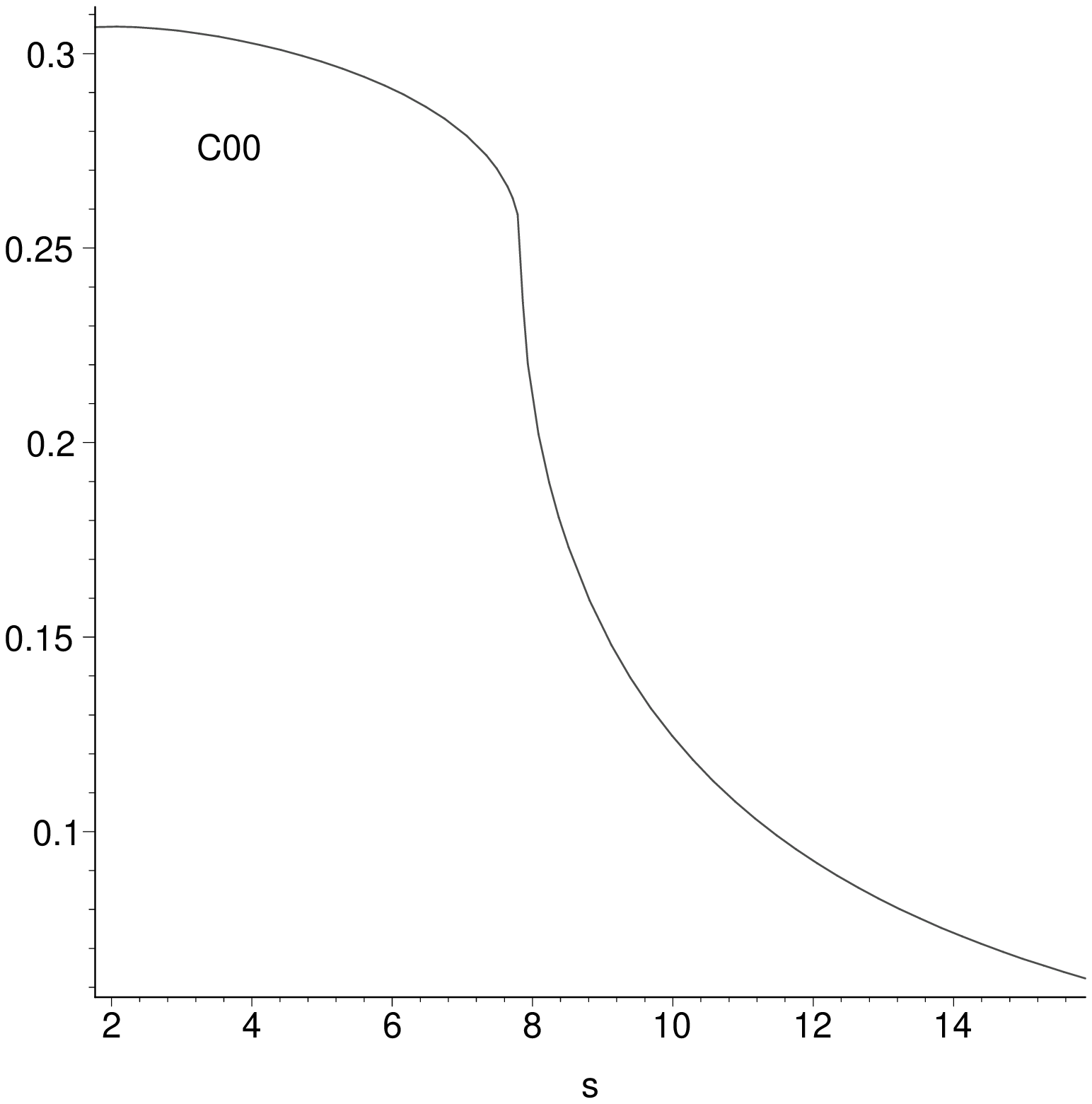}}}
\makebox[7cm]{\resizebox{7cm}{!}{\includegraphics[69,150][507,602]{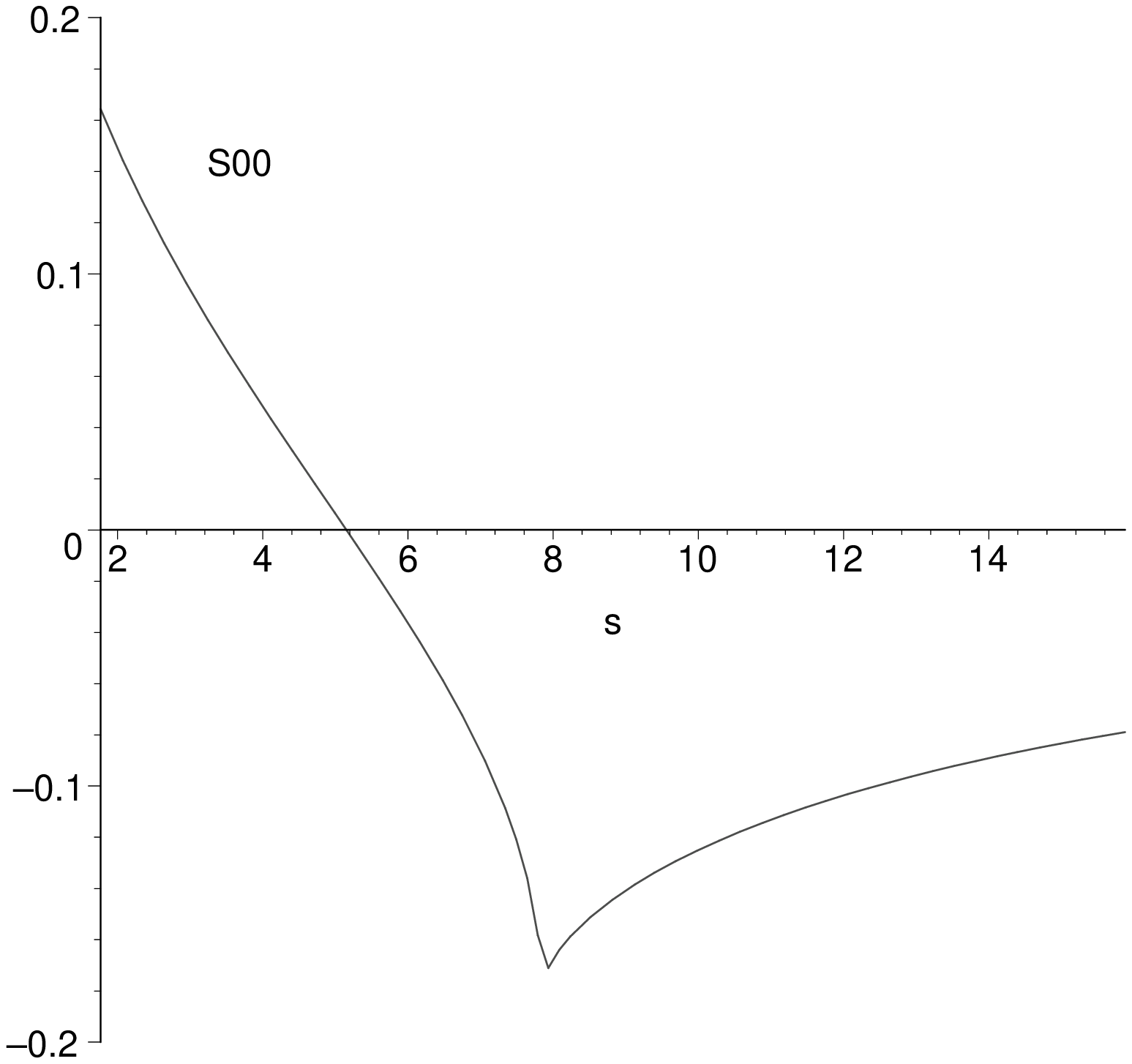}}}
\caption{${\cal C}$ and ${\cal S}$ parameters depending on the virtual gluon mass $q^2$ in the penguin model, for the decay 
$B^0 / \overline{B^0} \to K^0 \overline{K^0}$ \cite{Fleischer}.
\label{FigFleischerCS}}
\end{figure}

Of particular interest to us is the term $\Delta P$ in Eq. (\ref{AmpFleischer}), which represents the residual contribution of $c \overline{c}$ and
$u \overline{u}$ intermediate states in the QCD penguin diagrams \cite{BSS,Gerard;Hou}. This complex contribution depends on the effective 
(time-like) $q^2$ of the virtual gluon that mediates the $B \to K \overline{K}$ transition. An explicit expression for $\Delta P (q^2)$ is 
given in Ref. \cite{Fleischer}, and suggests that $| \Delta P |$ can be as large as $0.3$ for $\langle q^2 \rangle_{\rm eff} \approx \frac{1}{3} m^2_b$.
In the limit in which $\Delta P$ is neglected, the asymmetry parameters ${\cal C}$ and ${\cal S}$ for the decay $B^0 / \overline{B^0} \to K^0 \overline{K^0}$,
defined as
\begin{eqnarray}
{\cal C} = \frac{1 - | \lambda_{K^0 \overline{K^0}} |^2}{1 + | \lambda_{K^0 \overline{K^0}} |^2} & , & {\cal S} = 
\frac{2 \, {\rm Im} \lambda_{K^0 \overline{K^0}}}{1 + | \lambda_{K^0 \overline{K^0}} |^2}, \\
\lambda_{K^0 \overline{K^0}} = \frac{q}{p} \frac{A(\overline{B^0} \to K^0 \overline{K^0})}{A(\overline{B^0} \to K^0 \overline{K^0})} 
&, & \frac{q}{p} = e^{-2 i \beta} \nonumber
\end{eqnarray}
vanish identically: ${\cal C} = {\cal S} = 0$. Non-zero values of ${\cal C}$ and ${\cal S}$ are thus associated with effects of $c \overline{c}$
intermediate states, embedded in the function $\Delta P ( q^2)$. The behaviour of ${\cal C}$ and ${\cal S}$ as a function of $q^2$ is shown in
Fig. \ref{FigFleischerCS}, where we have used $\lambda_u = | \lambda_u | e^{-i \gamma} \cong 3.6 \times 10^{-3} e^{- i 60^{\circ}}$,
$\lambda_c = - | \lambda_c| \cong - 8.8 \times 10^{-3}$, $\lambda_t = - (\lambda_u + \lambda_c)$ and $ 2 \beta \cong 45^\circ$.

In the present note, we re-examine the amplitudes for the $B \to K \overline{K}$ channels, from the standpoint of inelastic final-state
interactions coupling the $K \overline{K}$ and $D \overline{D}$ systems, and suggest an alternative to the term $\Delta P \lambda_c$ in Eq.
(\ref{AmpFleischer}). The motivation comes from a recent analysis of the decay channels $B \to \pi \pi$ \cite{BSL}, in which we showed that
a mixing of the strong-interaction channels $\pi \pi$ and $D \overline{D}$ in the isospin $I=0$ state helps to resolve the puzzling observations
of large direct $CP$ violation in $B^0 / \overline{B^0} \to \pi^+ \pi^-$ \cite{BelleRecent} and an enhanced branching ratio for 
$B^0 / \overline{B^0} \to \pi^0 \pi^0$ \cite{BaBar;Belle}. There is a
natural explanation for these correlated effects in terms of final-state interactions in systems of physical hadrons, which respect the strong-interaction
conservation laws. Experiment \cite{BelleBrowder} shows that $D \overline{D}$ systems are produced with large branching ratio (i.e. large
decay amplitude, proportional to $\lambda_c$). Production of such systems in a real intermediate state, followed by a small mixing into the final state,
results in an important imaginary contribution to the decay amplitude. This gives rise to asymmetries, and can enhance an (initially suppressed) branching
ratio. The same physical idea will be implemented here, via the mixing of $D \overline{D}$ and
$K \overline{K}$ states with the same strong-interaction quantum numbers. Introduction of the $D \overline{D}$ as an intermediate state in the
decay $B \to K \overline{K}$ acts as a long-distance substitute for the $c \overline{c}$ contribution in the QCD penguin loop parametrized by
$\Delta P ( q^2)$. Since the discussion is on the level of physical hadronic channels, no reference is made to the uncertain variable $q^2$, which
appears in the short-distance contribution. Here we deal with the contributions to individual exclusive decays, of empirically-established \cite{BelleBrowder},
real intermediate states of physical hadrons. We obtain different, correlated results for asymmetries and branching ratios.

Mixing occurs between the following pairs of states, possessing definite strong-interaction quantum numbers.
\begin{eqnarray}
\left. \begin{array}{l} | K \overline{K} \rangle_+ \equiv \frac{1}{\sqrt{2}} ( K^0 \overline{K^0} + K^+ K^-) \\ 
| D \overline{D} \rangle_+ \equiv \frac{1}{\sqrt{2}} (D^+ D^- + D^0 \overline{D^0}) \end{array} \right\} & & 
I=0, \, G = +, \, J^{PC} = 0^{++} \label{StatesinBrackets} \\ 
\left. \begin{array}{l} | K \overline{K} \rangle_- \equiv \frac{1}{\sqrt{2}} ( K^0 \overline{K^0} - K^+ K^-) \\ 
| D \overline{D} \rangle_- \equiv \frac{1}{\sqrt{2}} (D^+ D^- - D^0 \overline{D^0}) \end{array} \right\} & & 
I=1, \, G = -, \, J^{PC} = 0^{++} \nonumber
\end{eqnarray}
As a consequence, the physical decay amplitudes of $\overline{B^0}$ to these states are \cite{BSL}
\begin{equation}
\left( \begin{array}{c} A^+_{K \overline{K}} \\ A^+_{D \overline{D}} \end{array} \right) = S_+^\frac{1}{2} 
\left( \begin{array}{c} \tilde{A}^+_{K \overline{K}} \\ \tilde{A}^+_{D \overline{D}} \end{array} \right), \, \,
\left( \begin{array}{c} A^-_{K \overline{K}} \\ A^-_{D \overline{D}} \end{array} \right) = S_-^\frac{1}{2} 
\left( \begin{array}{c} \tilde{A}^-_{K \overline{K}} \\ \tilde{A}^-_{D \overline{D}} \end{array} \right) \label{AmpConnections}
\end{equation}
Here $\tilde{A}^+_{K \overline{K}}$, $\tilde{A}^+_{D \overline{D}}$, $\tilde{A}^-_{K \overline{K}}$, $\tilde{A}^-_{D \overline{D}}$ denote
the ``bare" decay amplitudes, in the absence of final-state interactions. The mixing effects are contained in the rescattering matrices
$S_+^\frac{1}{2}$ and $S_-^\frac{1}{2}$, where $S_+$ and $S_-$ are the strong-interaction $S$ matrices between the pairs of states given in Eqs. 
(\ref{StatesinBrackets}). In the limit in which the elastic phase shifts for $(K \overline{K})_\pm \to (K \overline{K})_\pm$ and
$(D \overline{D})_\pm \to (D \overline{D})_\pm$ are neglected, the rescattering matrices $S_\pm^\frac{1}{2}$ reduce to the simple form
\cite{Donoghue}
\begin{equation}
S_+^\frac{1}{2} = \left( \begin{array}{cc} \cos \theta_+ & i \sin \theta_+ \\ i \sin \theta_+ & \cos \theta_+ \end{array} \right) , \, \, 
S_-^\frac{1}{2} = \left( \begin{array}{cc} \cos \theta_- & i \sin \theta_- \\ i \sin \theta_- & \cos \theta_- \end{array} \right) \label{Spmmatrices}
\end{equation}
where $\theta_+$ and $\theta_-$ are two phenomenological mixing angles. The physical decay amplitudes of $\overline{B^0}$ to the states 
$K^0 \overline{K^0}$ and $K^+ K^-$ can then be written as 
\begin{eqnarray}
A_{K^0 \overline{K^0}} & = & \frac{1}{2} \left[ \cos \theta_+ ( \tilde{A}_{K^0 \overline{K^0}}+ \tilde{A}_{K^+ K^-} ) + \cos \theta_- 
(\tilde{A}_{K^0 \overline{K^0}} - \tilde{A}_{K^+ K^-} ) \right] \nonumber \\
& & + \frac{1}{2} i \left[ \sin \theta_+ ( \tilde{A}_{D^+ D^-} + \tilde{A}_{D^0 \overline{D^0}} ) + \sin \theta_-
(\tilde{A}_{D^+ D^-} - \tilde{A}_{D^0 \overline{D^0}} \right] \\
A_{K^+ K^-} & = & \frac{1}{2} \left[ \cos \theta_+ ( \tilde{A}_{K^0 \overline{K^0}}+ \tilde{A}_{K^+ K^-} ) - \cos \theta_- 
(\tilde{A}_{K^0 \overline{K^0}} - \tilde{A}_{K^+ K^-} ) \right] \nonumber \\
& & + \frac{1}{2} i \left[ \sin \theta_+ ( \tilde{A}_{D^+ D^-} + \tilde{A}_{D^0 \overline{D^0}} ) - \sin \theta_-
(\tilde{A}_{D^+ D^-} - \tilde{A}_{D^0 \overline{D^0}} \right] \nonumber
\end{eqnarray}
For the charged decay mode $B^- \to K^- K^0$, involving the $I=1$ state $K^- K^0$, mixing with the state $D^- D^0$, with mixing angle $\theta_-$, implies
\begin{equation}
A_{K^- K^0} = \cos \theta_- \tilde{A}_{K^- K^0} + i \sin \theta_- \tilde{A}_{D^- D^0}
\end{equation}
For the bare amplitudes, we use the leading term of the QCD penguin model expressed by Eqs. (\ref{AmpFleischer}):
\begin{equation}
\tilde{A}_{K^0 \overline{K^0}} = \tilde{A}_{K^- K^0} = N_0 \lambda_t, \, \, \tilde{A}_{K^+ K^-} = 0.
\end{equation}
The bare amplitudes for $\overline{B^0} \to D^+ D^-$, $\overline{B^0} \to D^0 \overline{D^0}$ and $B^- \to D^- D^0$ are taken to be those in the 
Bauer-Stech-Wirbel factorization model \cite{BSW,Neubert;Stech}
\begin{equation}
\tilde{A}_{D^+ D^-} = \tilde{A}_{D^- D^0} = N^{\prime} \lambda_c, \, \, \tilde{A}_{D^0 \overline{D^0}} = 0
\end{equation}
where $N^{\prime} = 1.79$ is determined \cite{BSL} from the measured branching ratio ${\rm Br}(B^0 / \overline{B^0} \to D^+ D^-) \cong 2.5 \times 10^{-4}$
\cite{BelleBrowder}. With this input, the physical decay amplitudes, for small mixing angles ($\cos \theta_\pm \approx 1$,
$\sin \theta_\pm \approx \theta_\pm$) take the form
\begin{eqnarray}
A_{K^0 \overline{K^0}} & \cong & N_0 \lambda_t + i N^{\prime} \lambda_c \frac{1}{2} ( \theta_+ + \theta_-) \nonumber \\
A_{K^+ K^-} & \cong & i N^{\prime} \lambda_c \frac{1}{2} ( \theta_+ - \theta_-) \label{AmpinNlambda} \\
A_{K^- K^0} & \cong & N_0 \lambda_t + i N^{\prime} \lambda_c \theta_- \nonumber
\end{eqnarray}
The amplitudes satisfy the triangle relation (for all $\theta_\pm$)
\begin{equation}
A_{K^0 \overline{K^0}} - A_{K^+ K^-} = A_{K^- K^0} \label{TriangleRel}
\end{equation}
In what follows, we discuss the observable consequences of the amplitudes in Eqs. (\ref{AmpinNlambda}) and compare them with those obtained
from the short-distance penguin model expressed by Eqs. (\ref{AmpFleischer}).

\section{Branching Ratios (averaged over $B$ and $\overline{B}$)}

\begin{figure}
\center
\psfrag{Br00}[cc][][1.7]{${\rm Br}(B^0 / \overline{B^0} \to K^0 \overline{K^0})$}
\psfrag{Brpm}[bl][][1.7][90]{${\rm Br}(B^- / B^+ \to K^- K^0 / K^+ \overline{K^0})$}
\psfrag{1e\26106}[cc][][1.4]{$1 \times 10^{-6}$}
\psfrag{1.2e\26106}[cc][][1.4]{$1.2 \times 10^{-6}$}
\psfrag{1.4e\26106}[cc][][1.4]{$1.4 \times 10^{-6}$}
\psfrag{1.6e\26106}[cc][][1.4]{$1.6 \times 10^{-6}$}
\psfrag{1.8e\26106}[cc][][1.4]{$1.8 \times 10^{-6}$}
\psfrag{2e\26106}[cc][][1.4]{$2 \times 10^{-6}$}
\psfrag{2.2e\26106}[cc][][1.4]{$2.2 \times 10^{-6}$}
\makebox[9cm]{\resizebox{9cm}{!}{\includegraphics[49,150][531,602]{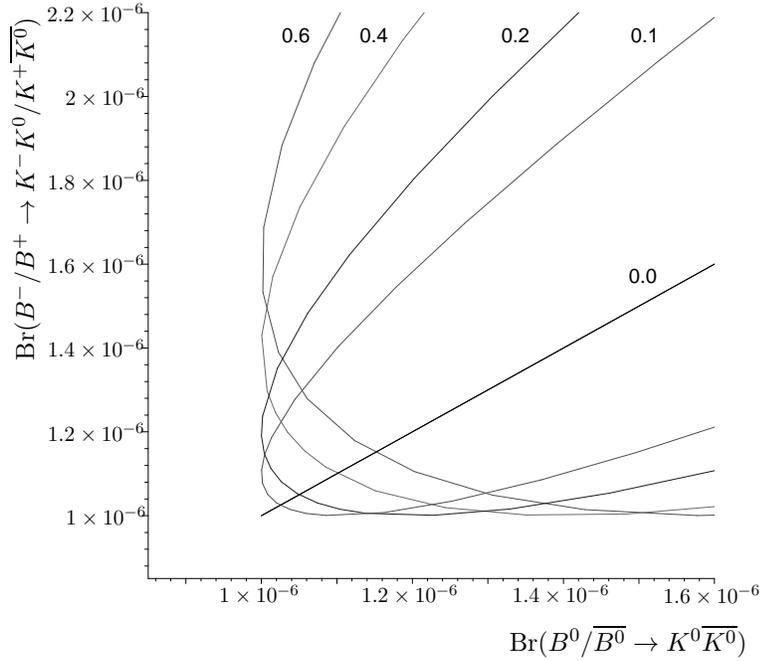}}}
\caption{${\rm Br}(B^- / B^+ \to K^- K^0 / K^+ \overline{K^0})$ as a function of ${\rm Br}(B^0 / \overline{B^0} \to K^0 \overline{K^0})$. 
The parameters $0.x$ describe the branching ratio of $B^0 / \overline{B^0} \to K^+ K^-$ in units of $10^{-6}$.
\label{FigBrofBr}}
\end{figure}

From Eqs. (\ref{AmpinNlambda}), we obtain the average branching ratios
\begin{eqnarray}
{\rm Br}(B^0 / \overline{B^0} \to K^0 \overline{K^0}) & = & | N_0 \lambda_t|^2 + N^{\prime 2} |\lambda_c|^2 \frac{1}{4} (\theta_+ + \theta_-)^2 \nonumber \\
{\rm Br}(B^0 / \overline{B^0} \to K^+ K^-) & = & N^{\prime 2} |\lambda_c|^2 \frac{1}{4} (\theta_+ - \theta_-)^2 \\
{\rm Br}(B^- / B^+ \to K^- K^0 / K^+ \overline{K^0}) & = & | N_0 \lambda_t|^2 + N^{\prime 2} |\lambda_c|^2 \theta_-^2 \nonumber
\end{eqnarray}
Notice that for different values of $\theta_+$ and $\theta_-$, the branching ratios of $B^0 / \overline{B^0} \to K^0 \overline{K^0}$ and
$B^- / B^+ \to K^- K^0 / K^+ \overline{K^0}$ are no longer equal. Futhermore, for $\theta_+ \neq \theta_-$, the decay rate of
$B^0 / \overline{B^0} \to K^+ K^-$ does not vanish, in contrast to the expectations from Eq. (\ref{AmpFleischer}). The correlation of the three branching
ratios is shown in Fig. \ref{FigBrofBr}, which gives the relation between the $K^0 \overline{K^0}$ and $K^- K^0 / K^+ \overline{K^0}$ rates for
given values of the $K^+ K^-$ branching ratio. The empirical limits on these decays are 
${\rm Br}(B^0 / \overline{B^0} \to K^0 \overline{K^0}) < 1.6 \times 10^{-6}$ \cite{Babar02}, 
${\rm Br}(B^0 / \overline{B^0} \to K^+ K^-) < 0.6 \times 10^{-6}$ \cite{BelleChao} and 
${\rm Br}(B^- / B^+ \to K^- K^0 / K^+ \overline{K^0}) < 2.2 \times 10^{-6}$ \cite{Babar03}. These data constrain the mixing angles $\theta_\pm$ to the
intervals $| \theta_+ + \theta_- | \leq 0.15$, $| \theta_+ - \theta_- |\leq 0.10$ and $|\theta_-| \leq 0.07$.

\section{${\cal C}$ and ${\cal S}$ Parameters for $B^0 / \overline{B^0} \to K^0 \overline{K^0}$}

\begin{figure}
\center
\psfrag{C00}[cc][][2]{${\cal C}$}
\psfrag{theta_p}[cc][][1.8]{$\theta$}
\psfrag{S00}[cc][][2]{${\cal S}$}
\makebox[7cm]{\resizebox{7cm}{!}{\includegraphics{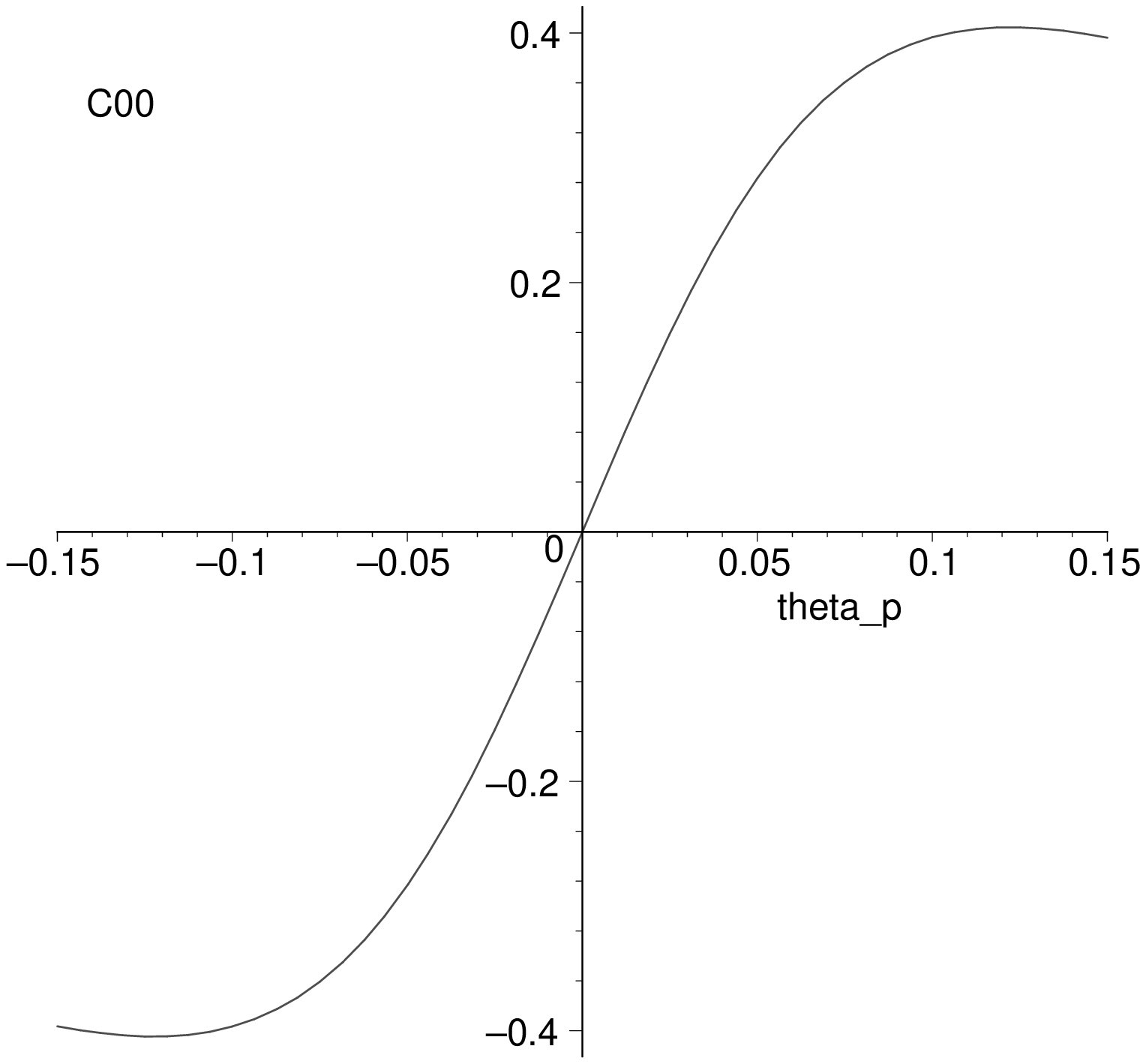}}}
\makebox[7cm]{\resizebox{7cm}{!}{\includegraphics[86,187][520,615]{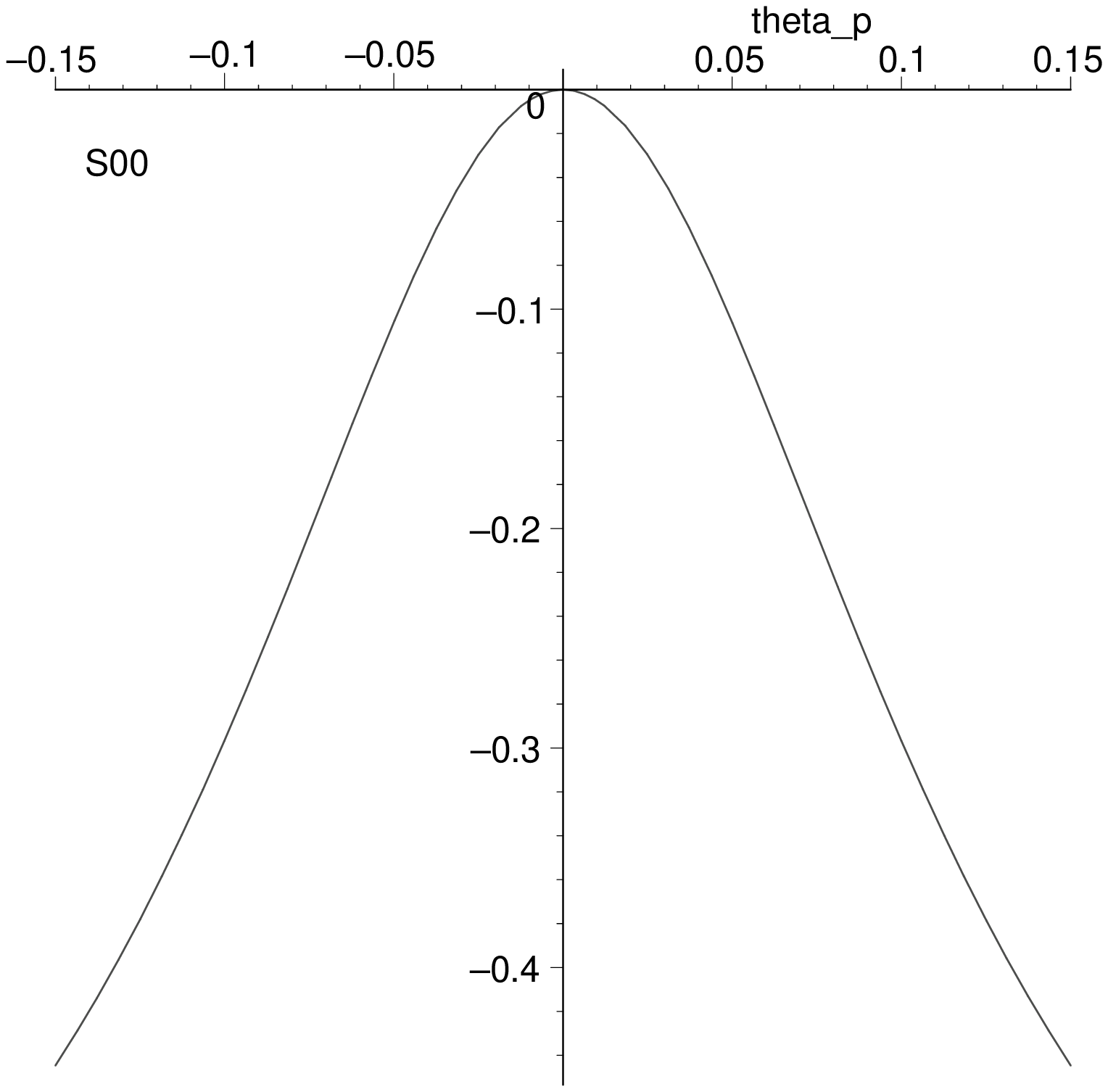}}}
\caption{${\cal C}$ and ${\cal S}$ parameters as functions of the mixing angle $\theta = \theta_+ + \theta_-$ in the decay 
$B^0 / \overline{B^0} \to K^0 \overline{K^0}$. \label{FigourCS00}}
\end{figure}

\begin{figure}
\center
\psfrag{C00}[cc][][2]{${\cal C}$}
\psfrag{S00}[cc][][2]{${\cal S}$}
\psfrag{tplus}[cc][][1.3]{$\theta = 0.15$}
\psfrag{tminus}[cc][][1.3]{$\theta = -0.15$}
\psfrag{qlow}[cc][][1.3]{$q^2 = 0.1 \, m^2_b$}
\psfrag{qup}[cc][][1.3]{$q^2 = 0.9 \, m^2_b$}
\makebox[9cm]{\resizebox{9cm}{!}{\includegraphics{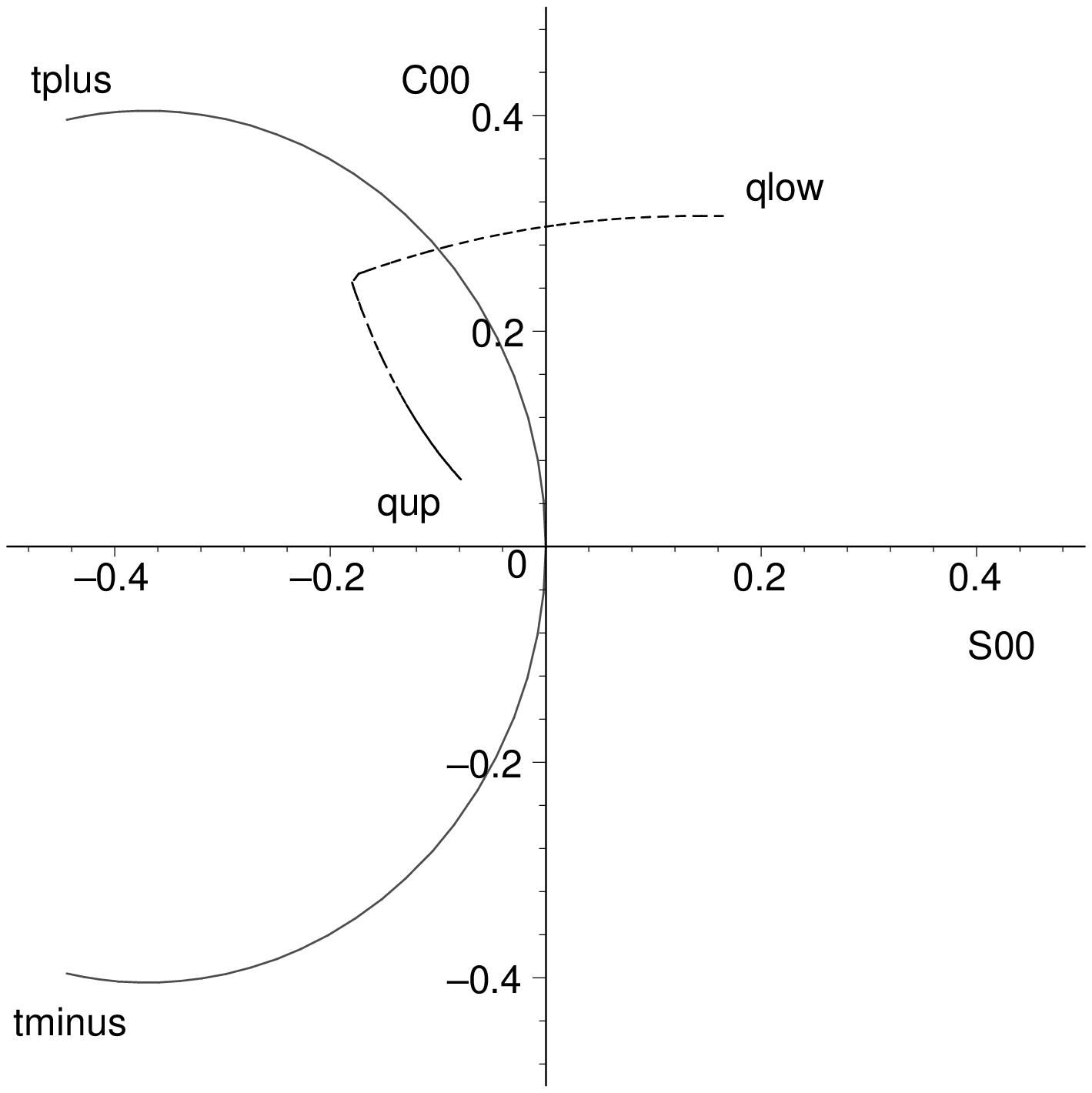}}}
\caption{${\cal C}$ as a function of ${\cal S}$ in the decay $B^0 / \overline{B^0} \to K^0 \overline{K^0}$.
Solid line: mixing model for parameters $-0.15 \leq \theta = \theta_+ + \theta_- \leq 0.15$, dashed line:
penguin model for $0.1 \, m^2_b \leq q^2 \leq 0.9 \, m^2_b$. \label{FigtrajCS}}
\end{figure}

The asymmetry parameters ${\cal C}$ and ${\cal S}$ for direct and indirect $CP$ violation respectively, calculated from the amplitudes in 
Eqs. (\ref{AmpinNlambda}), are shown in Fig. \ref{FigourCS00}.
For values of $| \theta_+ + \theta_-|$ of order $0.1$, the predicted values are $|{\cal C}| \approx 0.4$, ${\cal S} \approx -0.3$. To facilitate
comparison with the QCD penguin model, we show in Fig. \ref{FigtrajCS} the trajectory of allowed values in the ${\cal C} - {\cal S}$ plane. The
model with $D \overline{D} \leftrightarrow K \overline{K}$ mixing produces the trajectory drawn as a solid line, with points labelled by the value of
$\theta = \theta_+ + \theta_-$. The penguin model predicts the trajectory drawn as a dashed line, labelled by values of $q^2$, the virtual gluon mass. For the 
theoretically favoured region $\frac{1}{4} \leq \frac{q^2}{m^2_b} \leq \frac{1}{2}$ \cite{Gerard;Hou}, the short-distance model predicts
${\cal C} \approx 0.2$, ${\cal S} \approx -0.2$. While the two trajectories intersect, there is a large domain in which the predictions of the two
models are quite different. Since $A_{K^+ K^-}$ arises from the final-state interaction for $\theta_+ \neq \theta_-$, we have ${\cal C}_{K^+ K^-} \cong 0$
and ${\cal S}_{K^+ K^-} \cong - \sin 2 \beta \cong -0.7$.

\section{Effects of a Small Elastic Phase in the Rescattering Matrix}

\begin{figure}
\center
\psfrag{C00}[cc][][2]{${\cal C}$}
\psfrag{S00}[cc][][2]{${\cal S}$}
\psfrag{deltam}[cc][][1.8]{$\delta = -20^\circ$}
\psfrag{deltap}[cc][][1.8]{$\delta = +20^\circ$}
\makebox[7cm]{\resizebox{7cm}{!}{\includegraphics{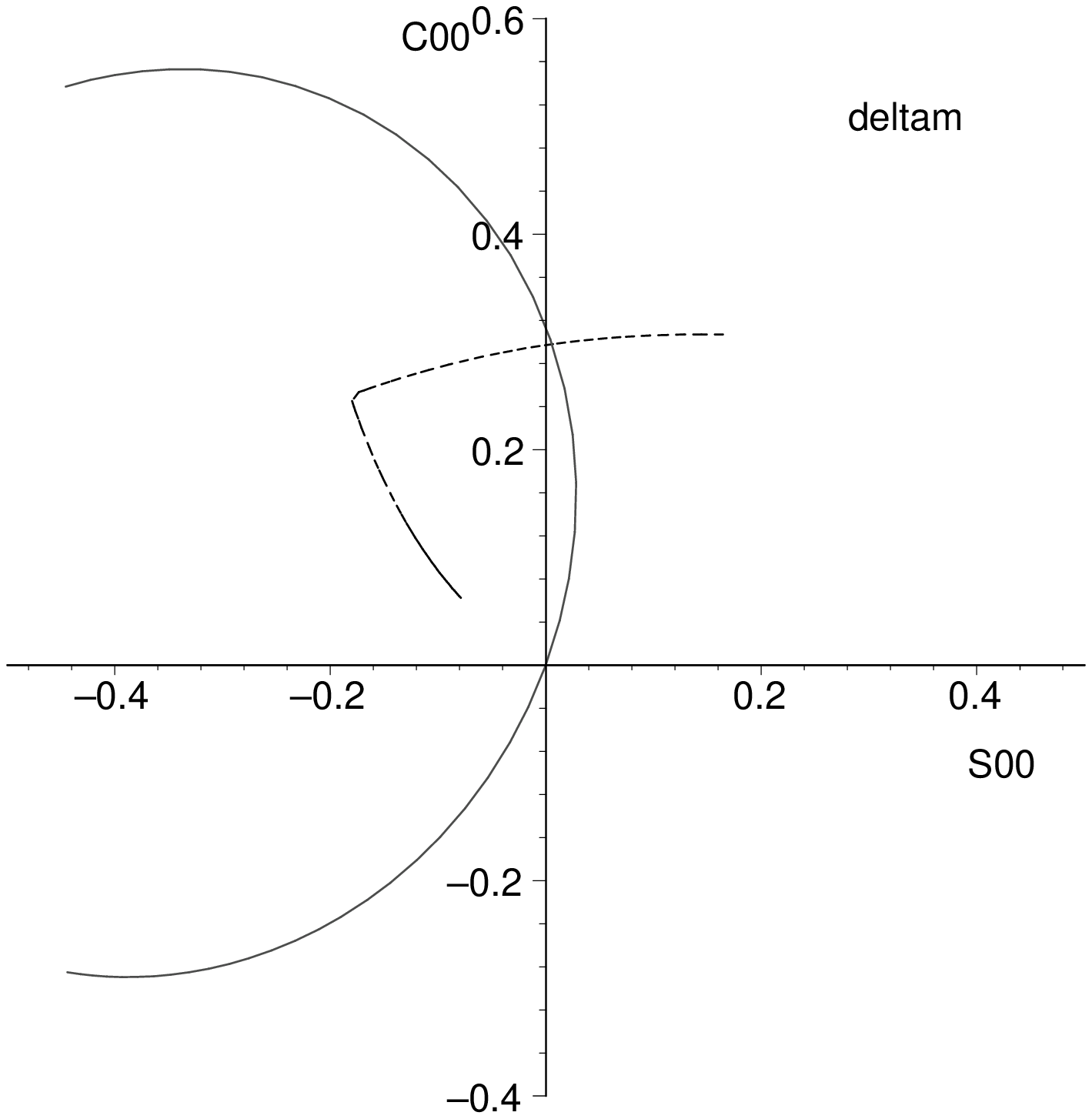}}}
\makebox[7cm]{\resizebox{7cm}{!}{\includegraphics{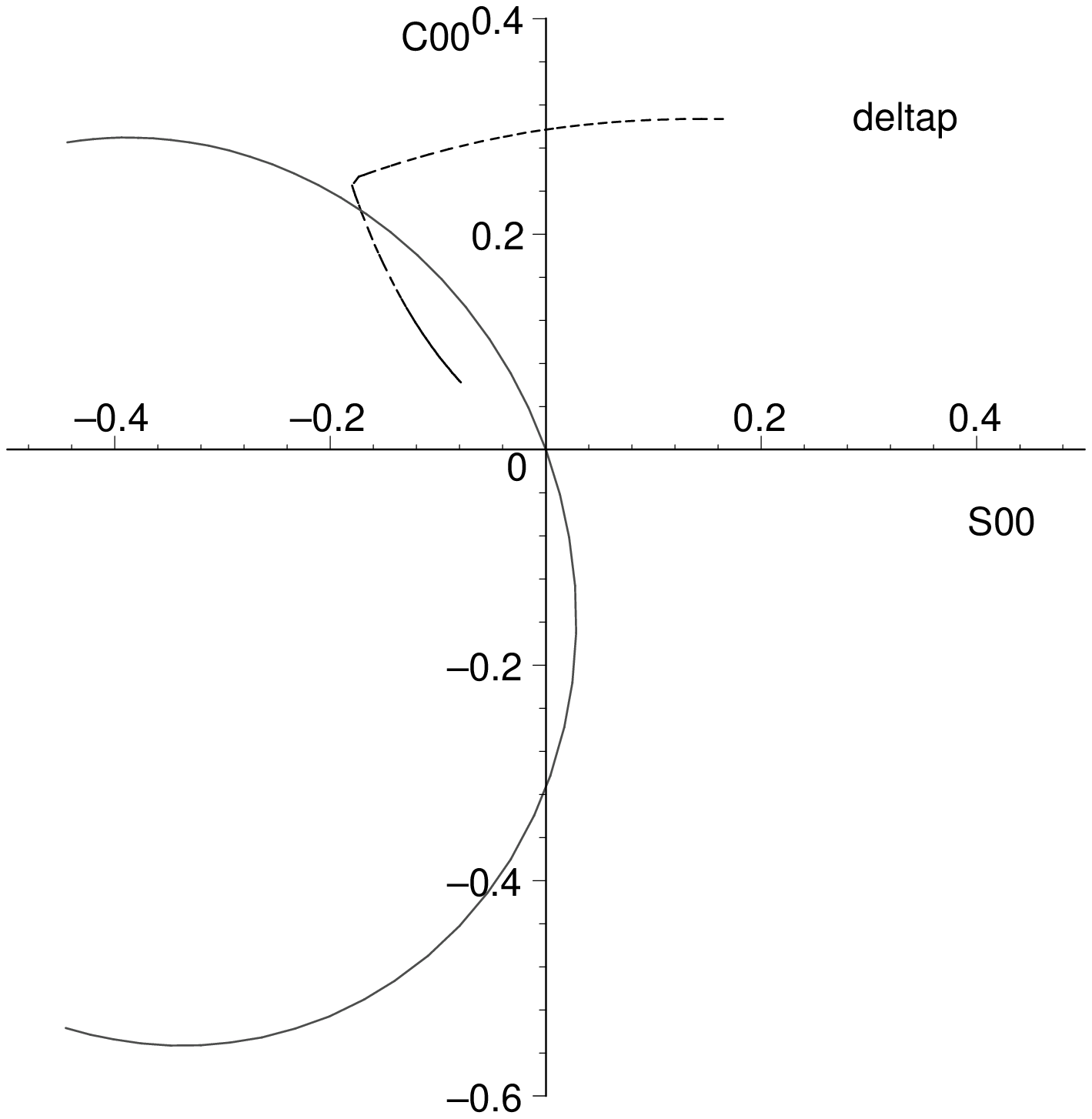}}}
\caption{ ${\cal C}$ as a function of ${\cal S}$ in the decay $B^0 / \overline{B^0} \to K^0 \overline{K^0}$.
Solid line: mixing model for parameters $-0.15 \leq \theta = \theta_+ + \theta_- \leq 0.15$ and two different $\delta$, dashed line:
penguin model for $0.1 \, m^2_b \leq q^2 \leq 0.9 \, m^2_b$.
\label{Figtrajdelta}}
\end{figure}

\begin{figure}
\center
\psfrag{Br00}[cc][][1.7]{${\rm Br}(B^0 / \overline{B^0} \to K^0 \overline{K^0})$}
\psfrag{Brpm}[bl][][1.7][90]{${\rm Br}(B^- / B^+ \to K^- K^0 / K^+ \overline{K^0})$}
\psfrag{1e\26106}[cc][][1.4]{$1 \times 10^{-6}$}
\psfrag{1.2e\26106}[cc][][1.4]{$1.2 \times 10^{-6}$}
\psfrag{1.4e\26106}[cc][][1.4]{$1.4 \times 10^{-6}$}
\psfrag{1.6e\26106}[cc][][1.4]{$1.6 \times 10^{-6}$}
\psfrag{1.8e\26106}[cc][][1.4]{$1.8 \times 10^{-6}$}
\psfrag{2e\26106}[cc][][1.4]{$2 \times 10^{-6}$}
\psfrag{2.2e\26106}[cc][][1.4]{$2.2 \times 10^{-6}$}
\makebox[9cm]{\resizebox{9cm}{!}{\includegraphics[49,150][531,602]{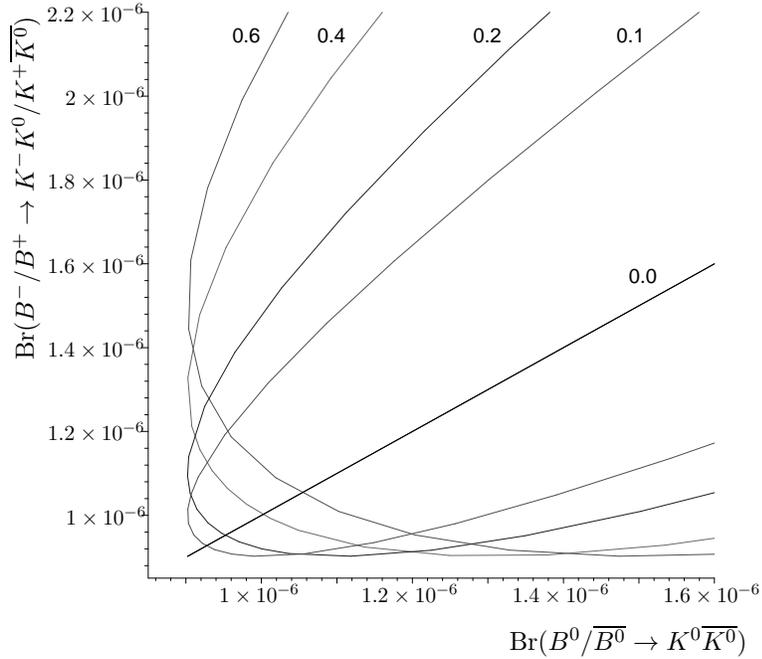}}}
\caption{${\rm Br}(B^- / B^+ \to K^- K^0 / K^+ \overline{K^0})$ as a function of ${\rm Br}(B^0 / \overline{B^0} \to K^0 \overline{K^0})$ for
a dynamical phase $\delta = \pm 20^{\circ}$. The parameters $0.x$ describe the branching ratio of $B^0 / \overline{B^0} \to K^+ K^-$ in units of $10^{-6}$.
\label{FigBrdelta}}
\end{figure}

The matrices $S_\pm^{\frac{1}{2}}$ that connect the bare and the physical amplitudes in Eq. (\ref{AmpConnections}) can have a more general form 
than that given in Eq. (\ref{Spmmatrices}), if the $S$ matrix is allowed to have non-zero elastic phases in the $K \overline{K}$ and
$D \overline{D}$ channels \cite{BSL}. For small phases, the net effect is to introduce a further phase factor $e^{i \delta}$ on
the second term in the $\overline{B^0} \to K^0 \overline{K^0}$ amplitude in Eq. (\ref{AmpinNlambda}). The effect of such a modification on the
trajectories in the ${\cal C} - {\cal S}$ plane is shown in Fig. \ref{Figtrajdelta}, for two choices $\delta = \pm 20^{\circ}$. The effect
on the average branching ratios is exhibited in Fig. \ref{FigBrdelta}.

\section{Comments on other Models for $B \to K \overline{K}$}

Predictions for $B \to K \overline{K}$ decays appear in various analyses dealing with $B \to PP$ transitions, where $P$
denotes a pseudoscalar meson. In Ref. \cite{CGRS}, an $SU(3)$ symmetry is assumed, and decay amplitudes are written as linear combinations of phenomenological
parameters associated with different quark topologies. These parameters are estimated from a fit to all $B \to PP$ channels. The model
gives equal rates for $K^0 \overline{K^0}$ and $K^- K^0$, and the $K^+ K^-$ rate is strongly suppressed, exactly as in the QCD penguin
amplitudes given in Eqs. (\ref{AmpFleischer}). A similar result occurs in the QCD factorization approach of Ref. \cite{Beneke;Neubert}, which
estimates an asymmetry parameter for $K^0 \overline{K^0}$ of ${\cal C} \approx 0.16$. In Ref. \cite{AKL}, the ${\cal C}$ and
${\cal S}$ parameters for $B^0 / \overline{B^0} \to K^0 \overline{K^0}$ are estimated to be ${\cal C} \approx 0.12$, ${\cal S} \approx -0.16$, 
similar to Ref. \cite{Fleischer}, using a model like 
that in Eqs. (\ref{AmpFleischer}). In Ref. \cite{CHY}, data on $B \to PP$ decays are fitted using elastic
scattering effects. The fits have very large elastic scattering phases. Then, they give a branching ratio for the $K^+ K^-$ channel of up to
$0.6 \times 10^{-6}$, a large asymmetry parameter ${\cal C}$ for the $K^0 \overline{K^0}$ channel, and also a ${\cal C}_{K^+ K^-}$
of the same sign. 
Smith \cite{Smith} has carried out an analysis of $B \to PP$ decays which discusses $SU(3)$ elastic scattering, and separately, effects
of $P \overline{P} \leftrightarrow D \overline{D}$ mixing using a parametrization different from ours. In the limit of retaining only the inelastic
mixing effect, his results for $B^0 / \overline{B^0} \to K^0 \overline{K^0}$ resemble ours, but the branching ratio $B^0 / \overline{B^0} \to K^+ K^-$
vanishes. In our approach, an enhanced rate for $K^+ K^-$ occurs as long as the mixing angles $\theta_+$ and $\theta_-$ for states of different isopin,
$G$-parity, are different.
Finally, there are papers discussing ``charming penguins", which attempt to parameterize charm-related,
long-distance aspects of the penguin diagram, but make no reference to strong-interaction eigenstates of physical hadrons in real intermediate states.
Results may be found in Ref. \cite{CFMPS} and the literature cited therein.

In conclusion, we note that current experiments are steadily approaching the level of sensitivity at which the decays $B \to K \overline{K}$ should
soon be measurable. One awaits with interest the pattern of branching ratios and asymmetries that these measurements will reveal. Our results here
show that this pattern can elucidate further important aspects of the underlying dynamics. The unexpected, recent oberservations of enhanced rates,
for $\pi^0 \pi^0$ \cite{BaBar;Belle}, $\pi^0 \rho^0$ \cite{BaBarRho,BelleDragie}, and $\eta \omega$ \cite{SmithRencontres} are striking consequences
of this dynamics \cite{BSL,BKS}.

\end{document}